\documentstyle[12pt]{article}

\renewcommand{\S}{{\cal S}}
\newcommand{\M}{{\cal M}}
\renewcommand{\le}{\partial_l}
\newcommand{\ri}{\partial_r}
\begin{document}

\begin{center}
{\LARGE\bf Master equation in the general gauge:}\\
{\Large\bf on the problem of infinite reducibility}
\end{center}
\begin{center}
{\bf G.A. Vilkovisky}
\end{center}
\begin{center}
Lebedev Physics Institute and Research Center in Physics,\\
Leninsky Prospect 53, Moscow 117924, Russia
\end{center}
\vspace{2cm}
\begin{abstract}
The master equation is quantized. This is an example of
quantization of a gauge theory with nilpotent generators.
No ghosts are needed for a generation of the gauge algebra.
The point about the nilpotent generators is that
one can't write down a single functional integral for this
theory. One has to write down a product of two coupled
functional integrals and take a square root. In the special
gauge where the gauge conditions are commuting, the functional
integrals decouple, and one recovers the known result.
\end{abstract}

\newpage

\begin{center}
\section*{\bf 1   Introduction}
\end{center}

$$ $$

When, in a gauge theory, the local generators of gauge
transformations are reducible, the quantization rules
are known only in the case where the reducibility has
a finite number of stages [1]. The case of infinite
reducibility is also of interest [1-4] but in this
case the usual methods of covariant quantization don't
work. An important subcase of infinite reducibility is the one
where the generators of gauge transformations are nilpotent.
Only this subcase is considered in the present paper, and,
furthermore, a solution to the quantization problem is
obtained only for a theory with equal numbers of boson
and fermion variables. This is the theory whose action
solves the master equation [1].

The master equation has two aspects. On the one hand, it
yields a universal formulation of any gauge theory (at least
finite-reducible). A given gauge theory is quantized by
building a dynamically equivalent solution of the master
equation. On the other hand, a solution of the master
equation itself is the action of a gauge theory with
nilpotent generators [1]. The quantization rules in this
theory have been known for one special class of gauges [1]
which thus far was sufficient for all practical purposes.
However, going beyond this class was notoriously difficult
because the master equation, being a theory with nilpotent
generators, has never been quantized properly. This problem
is solved below.

The solution is simple but has novel and unexpected features
which are probably common for all cases of nilpotent generators.
It is known that the higher the tower of reducibility the more
ghost fields are needed for unitarity
[1]. Therefore, naively, an infinite tower requires an infinite
number of ghosts. The correct answer is that, for a generation
of the gauge algebra with nilpotent generators, no ghosts are
needed at all. The solution is that, because of the nilpotency,
one can't write down the functional integral for a single copy
of the system. One has to double the system and
write down a product of two coupled functional integrals
for the doubled system. One is then to take a square root.
Only in the special class of gauges where
the gauge conditions are commuting, the functional integrals
decouple, and one recovers the previously known result.

\begin{center}
\section*{\bf 2   The master equation and the commuting gauge}
\end{center}

The master equation is the following equation:
\begin{equation}
(\S,\S)\equiv 2\frac{\ri\S}{\partial\phi^i}
\frac{\le\S}{\partial{\bar\phi}_i}=0
\end{equation}
for a boson function $\S(\phi,{\bar\phi})$ of fields $\phi^i$
and the "antifields" ${\bar\phi}_i$ having the statistics
opposite to the statistics of $\phi^i$. The notation $(\S,\S)$
refers to the operation of antibrackets [1].

Introducing the collective notation for the set of fields and
antifields:
\begin{equation}
\varphi^a =\phi^i , {\bar\phi}_i\;\;,\quad a=1,\ldots 2n
\end{equation}
\begin{equation}
\S(\varphi)=\S(\phi,{\bar\phi})\quad ,
\end{equation}
one can rewrite the master equation as follows [1]:
\begin{equation}
(\S,\S)\equiv \frac{\ri\S}{\partial\varphi^a}
\xi^{ab}\frac{\le\S}{\partial\varphi^b}=0\quad ,
\end{equation}
\begin{equation}
\xi^{ab}=
\left(
\begin{array}{cc}
{\displaystyle 0}&{\displaystyle\delta^i_k}\nonumber\\
{\displaystyle -\delta^k_i}&{\displaystyle 0}\nonumber\\
\end{array}
\right)\quad .
\end{equation}
Differentiating eq. (4) yields the Noether identities
\begin{equation}
\frac{\ri\S}{\partial\varphi^a}R^a_c=0\quad ,
\end{equation}
\begin{equation}
R^a_c=\xi^{ab}\frac{\le\ri\S}{\partial\varphi^b\partial\varphi^c}
\end{equation}
which show that every solution of the master equation is a gauge
action with the generators (7). Differentiating eq. (6) shows
that these generators are nilpotent on shell:
\begin{equation}
R^a_bR^b_c\:\biggl |_{\frac{\partial\S}{\partial\varphi}=0}=0\quad .
\end{equation}

The solution $\S$ of the master equation is assumed proper [1]:
\begin{equation}
\mbox{rank}\:\partial^2\S=\frac{2n}{2}
\end{equation}
in which case half of the variables $\varphi$ is redundant and
needs to be gauged away:
\begin{equation}
\chi_i(\varphi)=0\quad .
\end{equation}
For the special class of gauge conditions mentioned above, the gauge-fixing
functions $\chi_i(\varphi)$ are of the form [1]
\begin{equation}
\chi_i(\varphi)=\chi_i(\phi,{\bar\phi})={\bar\phi}_i-
\frac{\partial\Psi(\phi)}{\partial\phi^i}
\end{equation}
with an arbitrary fermion function $\Psi(\phi)$. Below, these
gauge conditions are referred to as $\chi^{\mbox{\scriptsize com}}(\varphi)$
where \mbox{"com"} stands for "commuting". Their canonical-invariant
characterization is
\begin{equation}
(\chi^{\mbox{\scriptsize com}}_i,\chi^{\mbox{\scriptsize com}}_j)=0\quad ,
\end{equation}
i.e., the gauge-fixing functions commute in the sense of antibrackets [1].
In this class of gauges, the functional integral generating the
Green's functions is of the form [1]
\begin{equation}
Z(J)=\int d\varphi\,\exp {\rm i}\Bigl(\S(\varphi)+\varphi^aJ_a\Bigr)\,
\delta\Bigl(\chi^{\mbox{\scriptsize com}}(\varphi)\Bigr)\quad .
\end{equation}
Here and below, the measure [1] is omitted but can always be
restored in the usual way.

No ghosts are needed in the commuting gauge. It is shown in the appendix
that the master equation always
admits a gauge of the form (11) at least locally. Apart from that, it has
always been a mystery, what makes this gauge distinguished but
attempts at a generalization didn't go through.

The answers come with quantizing the master equation. The procedure
below is an example of quantization of a gauge theory with nilpotent
generators.

\begin{center}
\section*{\bf 3   Quantizing the master equation}
\end{center}

For quantizing the master equation I shall build the master equation
for the master equation. Taking $\S(\varphi)$ for the original action
of a gauge field $\varphi^a$, I shall introduce the antifield
$\varphi^*_a$ and look for an action
\begin{equation}
\M(\varphi,\varphi^*)
\end{equation}
that would have $\S(\varphi)$ as its classical limit but would also
satisfy the master equation
\begin{equation}
\frac{\ri\M}{\partial\varphi^a}\frac{\le\M}{\partial\varphi^*_a}=0
\end{equation}
and be its proper solution:
\begin{equation}
\mbox{rank}\:\partial^2\M=\frac{4n}{2}\quad .
\end{equation}

The key point is that one needs no ghosts for building the solution
for $\M$. To see this, recall why ghosts are needed at all [1]. They
are needed for including the gauge generators in the hessian of the
action but, in the present case, the gauge generators of the original
$\S(\varphi)$ are already contained in its hessian, eq. (7). Therefore,
it should be possible to satisfy all the conditions for $\M$ without
introducing new fields. Indeed, here is the solution for $\M$:
\begin{equation}
\M(\varphi,\varphi^*)=\S(\varphi^a+\xi^{ab}\varphi^*_b)+
\S(\varphi^a-\xi^{ab}\varphi^*_b)\quad .
\end{equation}
It is easy to check that, with this expression, eq. (15) is satisfied
by virtue of the original master equation (4), and the rank condition
(16) is satisfied because the arguments of the two $\S$ 's in (17)
are independent. There remains to be inserted in (17) the overall 1/2
to satisfy the condition of the classical limit but {\it this is
precisely what should not be done}. It is another key point that the
classical limit should be kept doubled (see below).

In terms of the original fields and antifields, eq. (3), the solution
obtained is of the form
\begin{equation}
\M(\varphi,\varphi^*)=\S(\phi+{\bar\phi}^*\, ,\,{\bar\phi}-\phi^*)+
\S(\phi-{\bar\phi}^*\, ,\,{\bar\phi}+\phi^*)\quad ,
\end{equation}
\begin{equation}
\varphi=\phi,{\bar\phi}\;\quad ,\;\quad \varphi^*=\phi^*,{\bar\phi}^*
\end{equation}
where $\phi^*$ is the new antifield to the original field $\phi$ ,
and ${\bar\phi}^*$ is the new antifield to the original antifield
${\bar\phi}$ . The antifield to the antifield has, of course, the
statistics of the field.

The remaining procedure is standard. For the introduction of gauge
conditions one needs ghosts of the auxiliary sector [1]. One extends
$\M(\varphi,\varphi^*)$ in the usual way:
\begin{equation}
\M(\varphi,\varphi^*)+{\bar C}^*_i\pi^i=
\M_{\mbox{\scriptsize tot}}
(\varphi,\pi,{\bar C}\: ;\:\varphi^*,\pi^*,{\bar C}^*)
\end{equation}
where the bar over $C$ is needed only to hold to the conventional terminology
[1] since now there is only the ghost ${\bar C}$ ; there is no $C$ .
Then, by construction [1], the following functional integral:
\begin{eqnarray}
Z^2(J)&=&\int d\varphi d\pi d{\bar C} d\varphi^* d\pi^* d{\bar C}^*\,
\exp {\rm i}\Bigl(\M_{\mbox{\scriptsize tot}}
+2\varphi^aJ_a\Bigr)\nonumber\\
&\times&{}\delta\Bigl(\varphi^*-\frac{\partial\Psi}{\partial\varphi}\Bigr)
\delta\Bigl({\bar C}^*-\frac{\partial\Psi}{\partial{\bar C}}\Bigr)
\delta\Bigl(\pi^*-\frac{\partial\Psi}{\partial\pi}\Bigr)
\end{eqnarray}
does not depend on the choice of
\begin{equation}
\Psi=\Psi(\varphi,\pi,{\bar C})\quad .
\end{equation}

It is not a misprint that, in (21), $Z(J)$ appears squared, and the
source term $\varphi^aJ_a$ is doubled. The $Z(J)$ is the correct
generating functional for the master equation. Indeed, if one
decomposes $\varphi^a$ into the original $\phi^i,{\bar\phi}_i$ ,
and chooses
\begin{equation}
\Psi(\varphi,\pi,{\bar C})={\bar\phi}_i{\bar C}^i -\frac{1}{2}
\psi_1(\phi+{\bar C})+\frac{1}{2}\psi_2(\phi-{\bar C})
\end{equation}
with arbitrary functions $\psi_1,\psi_2$ , then upon the redesignations
\begin{eqnarray}
\varphi^a+\xi^{ab}\varphi^*_b&=&\varphi_1^a\; ,\\
\varphi^a-\xi^{ab}\varphi^*_b&=&\varphi_2^a
\end{eqnarray}
one obtains
\begin{equation}
Z^2(J)=\int d\varphi_1 d\varphi_2\,\exp {\rm i}\Bigl(\S(\varphi_1)+
\S(\varphi_2)+\varphi_1^aJ_a+\varphi_2^aJ_a\Bigr)\,
\delta\Bigl(\chi^{\mbox{\scriptsize com}}(\varphi_1)\Bigr)
\delta\Bigl(\chi^{\mbox{\scriptsize com}}(\varphi_2)\Bigr)\; .
\end{equation}
This is precisely the square of the functional integral (13) in
the commuting gauge!

\begin{center}
\section*{\bf 4   The master equation in the general gauge}
\end{center}

By setting in (22)
\begin{equation}
\Psi(\varphi,\pi,{\bar C})={\bar C}^i\chi_i(\varphi)
\end{equation}
with {\it arbitrary} $\chi_i(\varphi)$ one obtains the generating
functional $Z(J)$ in the general gauge:
\begin{equation}
Z^2(J)=\int d\varphi d{\bar C}\,\exp {\rm i}\biggl(
\S\Bigl(\varphi^a+\xi^{ab}{\bar C}^i\frac{\ri\chi_i}{\partial\varphi^b}\Bigr)+
\S\Bigl(\varphi^a-\xi^{ab}{\bar C}^i\frac{\ri\chi_i}{\partial\varphi^b}\Bigr)+
2\varphi^aJ_a\biggr)\,\delta\Bigl(\chi(\varphi)\Bigr)\;\; .
\end{equation}
In the general gauge, the $Z(J)$ itself cannot be written as a functional
integral. One has to write down a product 
of two coupled functional integrals and take a square root. The mystery 
of the commuting gauge is in the fact that, in this gauge, the functional 
integrals decouple. The explanation of the fact that, in the commuting
gauge, there are no ghosts is that the ghosts make another copy of the
original field. After the two copies decouple, each copy remains
without ghosts.   

Expanding in ${\bar C}$ in (28), one has
\begin{eqnarray}
Z^2(J)&=&\int d\varphi d{\bar C}\,\exp {\rm i}\biggl(
2\S(\varphi)+2\varphi^aJ_a
-{\bar C}^k\frac{\ri\chi_k}{\partial\varphi^e}\xi^{ea}
\frac{\le\ri\S}{\partial\varphi^a\partial\varphi^b}\xi^{bd}
\frac{\ri\chi_i}{\partial\varphi^d}{\bar C}^i
(-1)^{\varepsilon_i\varepsilon_b}\nonumber\\
&&\hspace{3cm}{}+O({\bar C}^4)\biggr)\,\delta\Bigl(\chi(\varphi)\Bigr)
\end{eqnarray}
with the obvious sign factor. It is seen that, owing to the doubling
in (17) and in the source term, the classical limit obtains correctly
but also, at the one-loop level, the ghost contribution doubles the
contribution of the gauge field. Indeed, the statistics of the ghosts
${\bar C}^i$ is opposite to the statistics of the redundant
variables (say, antifields) and is, therefore, the same as the
statistics of the independent variables (fields). Moreover, instead
of the usual ${\bar C}C$ in the quadratic term, one has now
${\bar C}{\bar C}$. The ghost determinant will, therefore, appear
to the power 1/2. Finally, the inverse ghost propagator contains
the hessian of the original gauge action which is the basic consequence
of the nilpotency of the gauge generators. There are, of course, also
the higher-order ghost couplings.

To summarize, the master equation possesses the baron M\"unchhausen's
gift of lifting itself up by the hair. The master equation (for the
master equation) in the commuting gauge yields the master equation
in the general gauge. Hence the notation $\M$ for the M\"unchhausen
master solution.

\begin{center}
\section*{\bf Acknowledgments}
\end{center}

This work is a product of the research program "Quantization,
Generalized BRS Cohomology, and Anomalies" organized by
the Erwin Schr\"odinger
International Institute for Mathematical Physics in autumn of 1998.
The author is grateful to the Erwin Schr\"odinger Institute and
the Technical University of Vienna for hospitality and support.
Special thanks to Anton Rebhan for his organizational efforts,
the physics discussions, and his enjoyable company in Vienna.
When in the black hole, the author is supported in part by the
Russian Foundation for Fundamental Research Grant 99-02-18107
and INTAS Grant 93-493-ext.

\newpage

\appendix
\makeatletter
\@addtoreset{equation}{section}
\makeatother
\renewcommand{\thesection}{Appendix.}
\renewcommand{\theequation}{A.\arabic{equation}}
\begin{center}
\section{\bf Existence of the commuting gauge}
\end{center}

The master equation always admits a commuting gauge at least in
some neighbourhood of the stationary orbit.
The proof is as follows.

Introduce the matrix inverse to (5):
\begin{equation}
\xi^{ab}\xi_{bc}=\delta^a_c
\end{equation}
and denote
\begin{equation}
\S_{ab}=\frac{\le\ri\S}{\partial\varphi^a\partial\varphi^b}
\:\biggl|_{\frac{\partial\S}{\partial\varphi}=0}\quad .
\end{equation}
The basic property of $\S_{ab}$ is nilpotency, eq. (8):
\begin{equation}
\S_{ab}\xi^{bc}\S_{cd}=0\quad .
\end{equation}
Since $\S$ is the proper solution, eq. (9), the $\S_{ab}$ has
a symmetric invertible minor of dimension $n\times n$ :
\begin{equation}
\mbox{rank}\:\Bigl(\delta^a_i\S_{ab}\delta^b_j\Bigr)=\frac{2n}{2}\; ,
\qquad i,j,k=1,\ldots n\; .
\end{equation}
Introduce the matrix inverse to this minor:
\begin{equation}
\Bigl(\delta^a_k\S_{ab}\delta^b_i\Bigr)G^{ij}=\delta^j_k\;\; ,
\end{equation}
and, with its aid, define the following quantities:
\begin{eqnarray}
\phi^i&=&\Bigl(G^{ij}\delta^b_j\S_{ba}\Bigr)\varphi^a\quad ,\\
{\bar\phi}_i&=&\delta^b_i\Bigl(-\xi_{ba}+
\frac{1}{2}\xi_{bc}\delta^c_kG^{kj}\delta^d_j\S_{da}\Bigr)\varphi^a\quad .
\end{eqnarray}

Eqs. (A.6),(A.7) define a linear transformation from the variables
$\varphi^a$ to the new variables $\phi^i,{\bar\phi}_i$ . This
transformation is invertible and is, moreover, a canonical transformation.
Indeed, one can check on the basis of the nilpotency relation (A.3)
that the inverse transformation is of the form
\begin{equation}
\varphi^a=\Bigl(\delta^a_b-\frac{1}{2}\xi^{ae}\S_{ec}\delta^c_j
G^{jk}\delta^d_k\xi_{db}\Bigr)\delta^b_i\,\phi^i-
\Bigl(\xi^{ac}\S_{cb}\delta^b_jG^{ji}\Bigr)\,{\bar\phi}_i\;\; ,
\end{equation}
and, in terms of the antibrackets
\begin{equation}
(A,B)\stackrel{\mbox{\scriptsize def}}{=}\frac{\ri A}{\partial\varphi^a}
\xi^{ab}\frac{\le B}{\partial\varphi^b}\quad ,
\end{equation}
the following relations hold:
\begin{equation}
(\phi^i,\phi^j)=0\quad ,\quad (\phi^i,{\bar\phi}_j)=\delta^i_j
\quad ,\quad ({\bar\phi}_i,{\bar\phi}_j)=0\quad .
\end{equation}

The canonical transformation above
settles the question. Indeed, a calculation of the hessian
of $\S$ in the new variables yields the following result:
\begin{equation}
\frac{\le\ri\S}{\partial\phi^i\partial\phi^j}
\:\biggl|_{\frac{\partial\S}{\partial\varphi}=0}=
\delta^a_i\S_{ab}\delta^b_j\quad ,\quad
\frac{\le\ri\S}{\partial{\bar\phi}_i\partial\phi^j}
\:\biggl|_{\frac{\partial\S}{\partial\varphi}=0}=0\quad ,\quad
\frac{\le\ri\S}{\partial{\bar\phi}_i\partial{\bar\phi}_j}
\:\biggl|_{\frac{\partial\S}{\partial\varphi}=0}=0\quad .
\end{equation}
By (7), the admissibility condition for a gauge-fixing function
$\chi_i$ is of the form
\begin{equation}
\mbox{rank}\:\frac{\ri\chi_i}{\partial\varphi^a}\xi^{ab}
\frac{\le\ri\S}{\partial\varphi^b\partial\varphi^c}
\:\biggl|_{\frac{\partial\S}{\partial\varphi}=0}=n\quad .
\end{equation}
When written in the variables $\phi,{\bar\phi}$ , with eqs. 
(A.11) and (A.4) taken into account, this condition becomes
\begin{equation}
\mbox{rank}\:\frac{\ri\chi_i(\phi,{\bar\phi})}{\partial{\bar\phi}_j}
\:\biggl|_{\frac{\partial\S}{\partial\varphi}=0}=n\quad .
\end{equation}
It follows that any gauge condition solvable with respect to
${\bar\phi}_i$ is admissible. This includes both commuting and
noncommuting gauges for which
\begin{equation}
\chi_i={\bar\phi}_i-f_i(\phi)
\end{equation}
with any functions $f_i(\phi)$ .

\begin{center}
\section*{\bf References}
\end{center}

\begin{enumerate}
\item I.A. Batalin and G.A. Vilkovisky, Phys. Lett. 102B (1981) 27;
Phys. Rev. D28 (1983) 2567; Phys. Rev. D30 (1984) 508.
\item R. Kallosh, W. Troost, and A. Van Proeyen, Phys. Lett. 212B
(1988) 428.
\item A. Van Proeyen, Talk at the "Strings and Symmetries 1991" conference
in Stony Brook, hep-th/9109036.
\item E. Bergshoeff, R. Kallosh, T. Ortin, and G. Papadopoulos,
Nucl. Phys. B502 (1997) 149.
\end{enumerate}
\end{document}